\documentclass{iau}
\usepackage{graphicx}

\title[The MiMeS Survey of Magnetism in Massive Stars] 
{The magnetic characteristics of Galactic OB stars from the MiMeS survey of magnetism in massive stars}

\author[Gregg A. Wade, J. Grunhut \& The MiMeS Collaboration]   
{G.A. Wade$^1$, J. Grunhut$^2$, E. Alecian$^{3,4}$, C. Neiner$^4$, M. Auri\`ere$^5$, D.A. Bohlender$^6$, A. David-Uraz$^1$, C. Folsom$^5$, H.F. Henrichs$^{7}$, O. Kochukhov$^8$, S. Mathis$^{9,4}$, S. Owocki$^{10}$, V. Petit$^{10}$\\ \and the MiMeS Collaboration}

\affiliation{$^1$RMC, Canada, $^2$ESO, Germany, $^3$IPAG, France, $^4$LESIA, France, $^5$IRAP, France, $^6$NRC, Canada, $^{7}$University of Amsterdam, Netherlands, $^8$Uppsala University, Sweden, \\$^9$CEA, France,$^{10}$University of Delaware, USA}

\pubyear{2008}
\volume{xxx}  
\pagerange{119--126}
\jname{Title of your IAU Symposium}
\editors{A.C. Editor, B.D. Editor \& C.E. Editor, eds.}
\begin{document}

\maketitle

\begin{abstract}
The Magnetism in Massive Stars (MiMeS) project represents the largest systematic survey of stellar magnetism ever undertaken. Based on a sample of over 550 Galactic B and O-type stars, the MiMeS project has derived the basic characteristics of magnetism in hot, massive stars. Herein we report preliminary results.
\end{abstract}

\firstsection 
\section{Introduction}

Near the main sequence, classical observational tracers of dynamo activity fade and disappear amongst stars of spectral type F, at roughly the conditions predicting the disappearance of energetically-important envelope convection. As an expected consequence, the magnetic fields of hotter stars differ significantly from those of cooler FGKM stars. They are detected in only a small fraction of stars, they are structurally much simpler, and frequently much stronger, than the fields of cool stars (e.g. Donati \& Landstreet 2009). They exhibit stability of their large-scale and smaller-scale structures on timescales of decades (e.g. Silvester et al. 2013). Most remarkably, their characteristics show no clear correlations with basic stellar properties such as age, mass or rotation.

These puzzling characteristics support a fundamentally different field origin than that of cool stars: that the observed fields are not currently generated by dynamos, but rather that they are {\em fossil fields}; i.e. remnants of field accumulated or generated during earlier phases of stellar evolution (e.g. Mestel 1999). 

The primary aim of the Magnetism in Massive Stars (MiMeS) project is to understand the origin and impact of magnetic fields in hot, massive stars, both from the observational and theoretical perspectives. In this paper we briefly report results from the analysis of the OB stars observed within the MiMeS survey. 

\section{The MiMeS survey}

The MiMeS 'survey component' (SC) was developed to provide critical missing information about the incidence and statistical  properties of organized magnetic fields in a large sample of massive stars. Over 4800 high precision (median SNR$\sim$800 per pixel), high resolution ($R\sim 65\, 000$) broad-bandpass ($364-1000$~nm) circularly polarized (Stokes $V$) spectra were acquired for approximately 550 OB stars ranging in spectral type from B9 to O5, in $V$ magnitude from 0.1 to 13.6, and in luminosity class from V to Ia. Data were acquired using the ESPaDOnS (CFHT), Narval (TBL) and HARPSpol (ESO3.6) spectropolarimeters.  Observations were obtained both in the context of competitively-allocated Large Programs (PIs Alecian (ESO), Neiner (TBL), Wade (CFHT)) and PI programs. Reduced, continuum-normalized Stokes $I$, $V$ and diagnostic null ($N$) spectra were primarily analyzed using Least-Squares Deconvolution (in particular, the iLSD implementation of Kochukhov et al. 2010). Magnetic diagnosis was performed in several ways. First, standard $\chi^2$ analysis (e.g. Donati et al. 1997) was performed to quantitatively assess the detection of any signal in the LSD mean $V$ or $N$ profile. Secondly, the longitudinal field was inferred from measurement of the first-order moment of the $V$ and $N$ profiles. Finally, the Bayesian statistics-based method of Petit \& Wade (2012) was applied to evaluate the odds ratio and the implied probability distribution of the surface magnetic field strength (again, using both the $V$ and $N$ profiles), under the assumption of a dipolar surface field configuration.

\section{Data quality and quality control}

\begin{figure}
\begin{center}
\includegraphics[width=6.25cm]{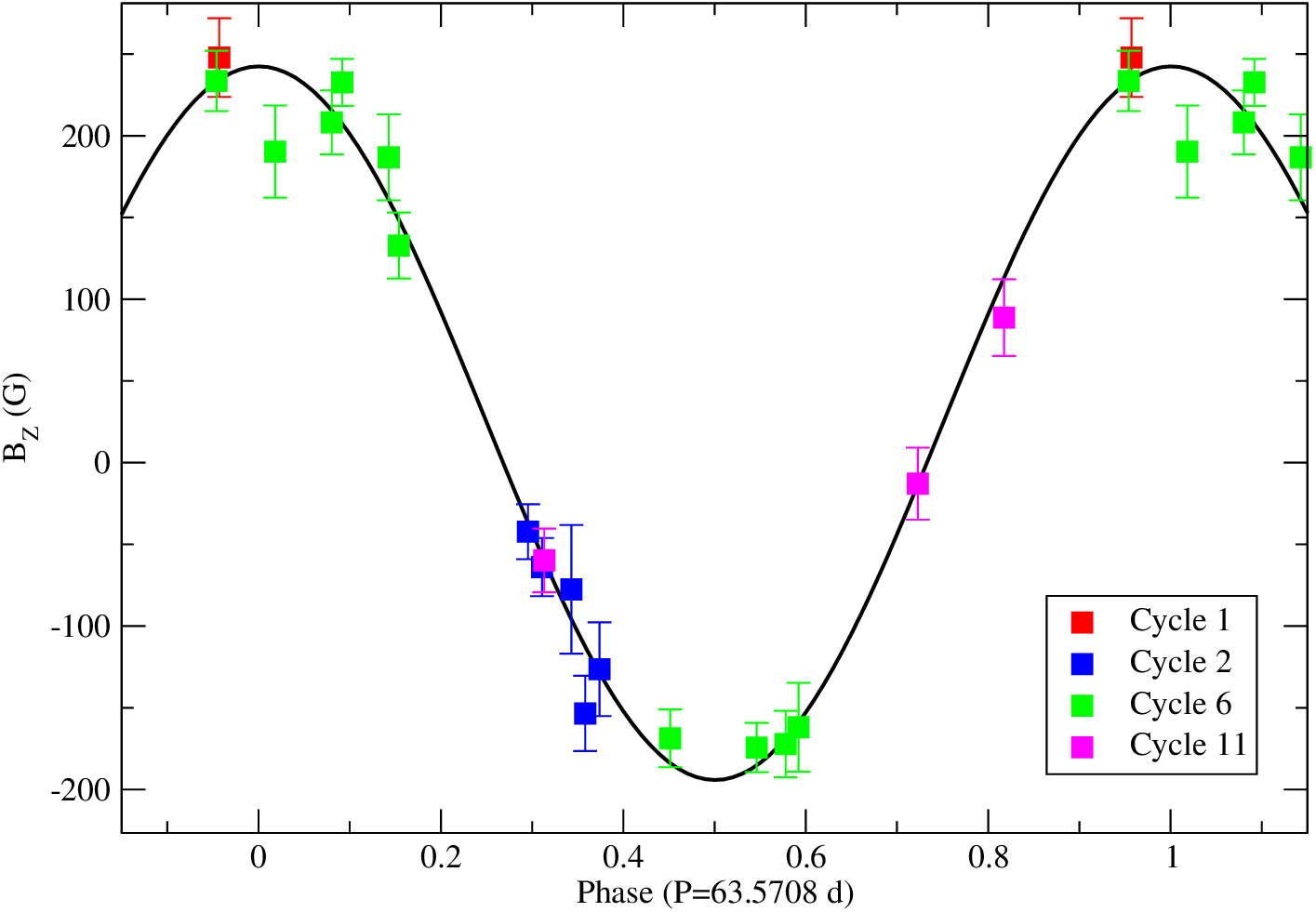}\hspace{0.5cm}\includegraphics[width=6.25cm]{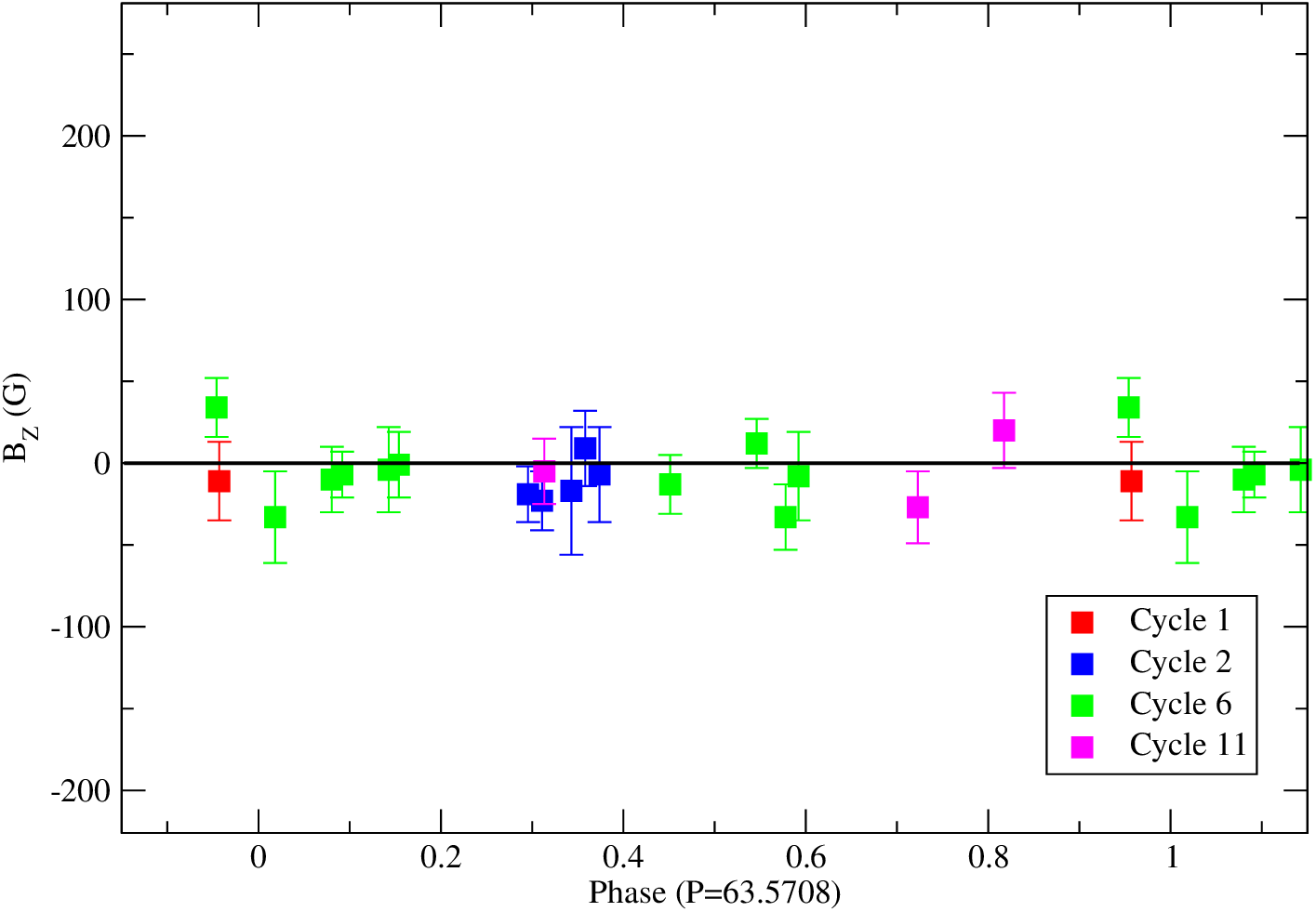}
\caption{Longitudinal magnetic field of the magnetic O9 subgiant star HD 57682, measured from Stokes $V$ (left) and $N$ (right) spectra acquired with ESPaDOnS and folded according to the rotational period of 63.6~d. Different colours indicate different epochs of observation from December 2008 to December 2010. HD 57682 was discovered to be magnetic within the context of the survey (Grunhut et al. 2009); the coherence of the data achieved through continued monitoring firmly establishes the existence and characteristics of the magnetic field, and illustrates the accuracy and precision of MiMeS magnetometry. }
\end{center}
\end{figure}

The precision of magnetic diagnoses carried out using high resolution Stokes $V$ spectra is a function of signal-to-noise ratio, spectral type and line width. For each target observed within the LPs, the exposure time was normally computed so as to reach a particular sensitivity threshold (0.1, 0.3, 0.5, 1.0 kG estimated surface dipole field) accounting for its particular apparent magnitude, spectral type and line width. Typically, the threshold identified for each star was that which allowed a total exposure time shorter than 2 hours. PI data added to the survey were generally of similar or better quality, due to the nature of many of the PI observing program goals.

In addition to the survey, a major undertaking of the MiMeS project is the detailed analysis of previously known and newly-detected magnetic massive stars as part of the so-called 'targeted component' (TC). Datasets obtained for these stars allow frequent assessment of the reliability of the measurements, which is key to ensuring that the results of the SC are trustworthy. MiMeS quality control focuses on the reproducibility of our measurements and verification of their associated uncertainties (both from the Zeeman signatures and corresponding longitudinal field measurements). This is accomplished in 3 basic ways. First, we confirm that the Stokes $V$ profile and longitudinal field are reproduced within the formal uncertainties in observations of individual magnetic targets acquired at similar rotational phases. Second, we employ the $N$ spectrum to test the instrumental systems for spurious contributions to the polarization. Finally, we examine the statistics of the observed SC sample with no detectable Zeeman signatures in a variety of ways, to ensure that the distribution can be ascribed to noise consistent with the expected observational uncertainties. 

\section{Results}

Of the $\sim 550$ stars observed within the context of the MiMeS project, approximately 65 show evidence for magnetic fields. Of the detected targets, about 30 were firmly identified as magnetic stars prior to the survey (i.e. these represent the "targeted component"). The statistics that follow are computed ignoring these 30 stars, i.e. based only on the 'blind' MiMeS survey results for $\sim 525$ stars.

The bulk incidence (i.e. the total number of previously-unknown magnetic stars in the sample relative to the total sample) is $7\pm 1$\%\ (n.b. all incidence uncertainties are computed from counting statistics). Of the approximately 430 B-type stars in the sample previously unknown to host magnetic fields, 32 are found to be magnetic for an incidence of $7\pm 1$\%. Of the approximately 90 O-type stars in the sample previously unknown to host magnetic fields, 6 are found to be magnetic for an incidence of $7\pm 3$\%. The incidence as a function of spectral type is illustrated in Fig. 2.

The detected magnetic stars (e.g. Grunhut et al. 2009, Alecian et al. 2011, Wade et al. 2012, Briquet et al. 2013) exhibit periodically variable longitudinal fields (with periods in the range of 0.5d to many years) corresponding to organized magnetic fields with important dipole components. The polar strengths of the dipoles range from several hundred G up to 20 kG. The general magnetic characteristics of detected B-type stars and O-type stars are very similar.

Thus we conclude that the incidences and characteristics of large-scale magnetic fields in B and O type stars are indistinguishable, and qualitatively identical to those of intermediate mass stars with spectral types $\sim$F0 to A0 on the main sequence. {\bf The MiMeS survey therefore establishes that the basic physical characteristics of magnetism in stellar radiative zones remains unchanged across more than 1.5 decades of stellar mass, from spectral types F0 ($\sim 1.5~M_\odot$) to O4 ($\sim 50~M_\odot$).}

\subsection{Subsample results}


\noindent {\bf Open clusters:}\ Through the HARPSpol LP, we have acquired magnetic observations of the complete populations of OB stars in 7 Galactic open clusters. In four of the observed clusters, no magnetic stars are detected (although we note that each of these clusters contains only 7 or 8 OB stars). The remaining clusters show incidences of 10-20\%, based on samples of 10-60 stars. This can be contrasted with the incidence of $\sim 30\%$ inferred by Petit et al. (2008) in the ONC. The inferred magnetic incidence vs. cluster age is illustrated in Fig. 2 (right panel).

\noindent {\bf Of?p stars:}\ We have observed all known Galactic stars of the Of?p class, and detected or confirmed the presence of magnetic field in every star of the sample (see Wade et al. 2012 and references therein). We therefore conclude that the {\bf Of?p stars represent a magnetic class, and that their peculiar spectral properties are likely a consequence of the interaction of their winds and magnetic fields} (e.g. Sundqvist et al. 2012).

\noindent{\bf Pulsating $\beta$~Cep/SPB stars:}\ We have observed 104 pulsating $\beta$~Cep and SPB stars. We infer a magnetic incidence in this subsample fully consistent with the larger sample (e.g. Shultz et al. 2012), indicating that pulsating {\bf $\beta$~Cep and SPB stars are not preferentially magnetic.} 

\noindent{\bf Classical Be stars:}\ We have observed a sample of 98 classical Be stars, and failed to obtain any detections despite a magnetic sensitivity similar to that of the larger sample. Based on the magnetic incidence measured for the non-Be B-type stars, we would have expected to detect $10\pm 2$ magnetic stars amongst the Be stars. Note that magnetic fields have been detected in other emission-line stars, e.g. Herbig Be stars. Hence we conclude that the lack of detected magnetic classical Be stars is a significant result, indicating that {\bf decretion Keplerian Be discs are not of magnetic origin} (Neiner et al. 2012).

\begin{figure}
\begin{center}
\includegraphics[width=6.25cm]{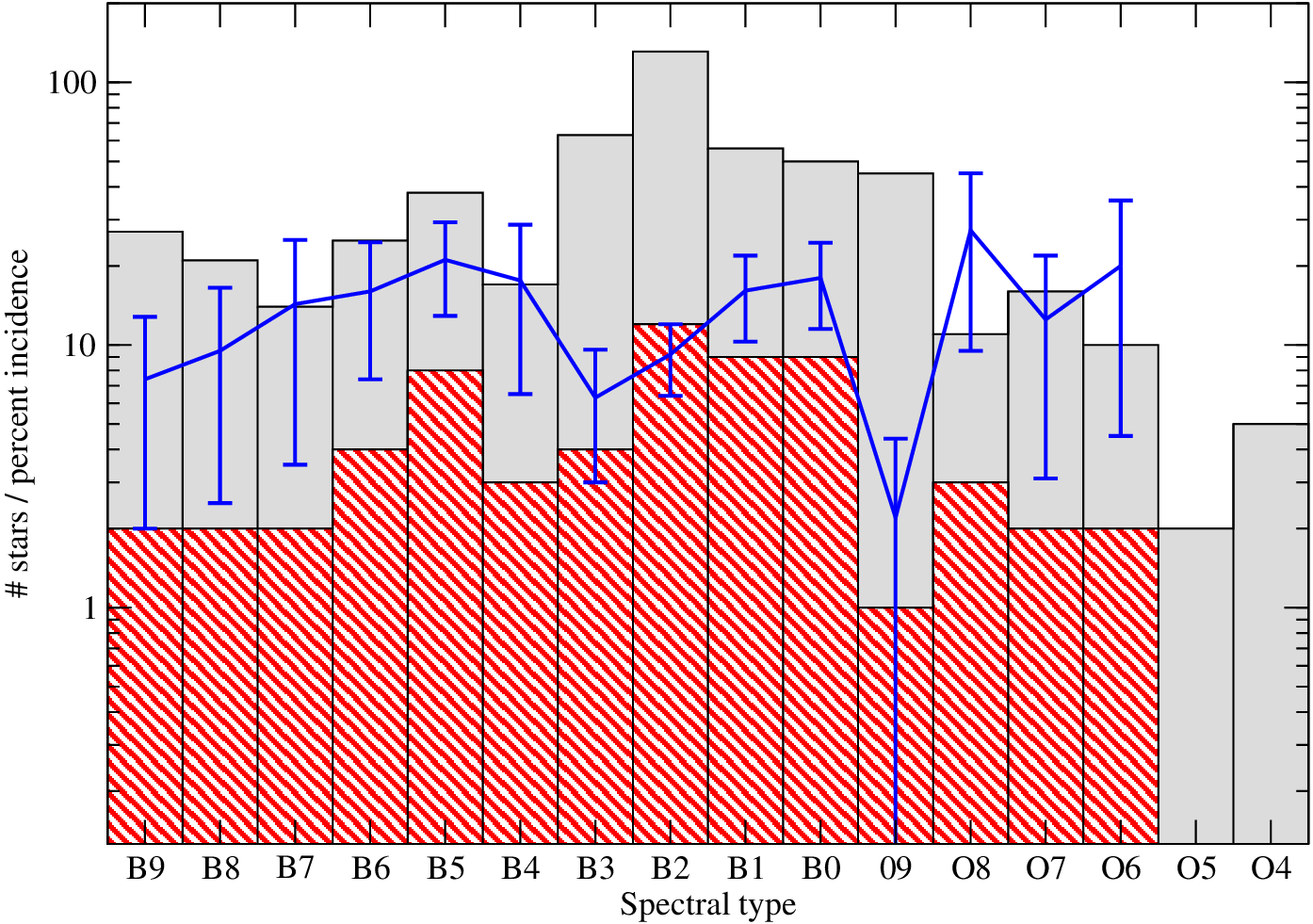}\hspace{0.5cm}\includegraphics[width=6.25cm]{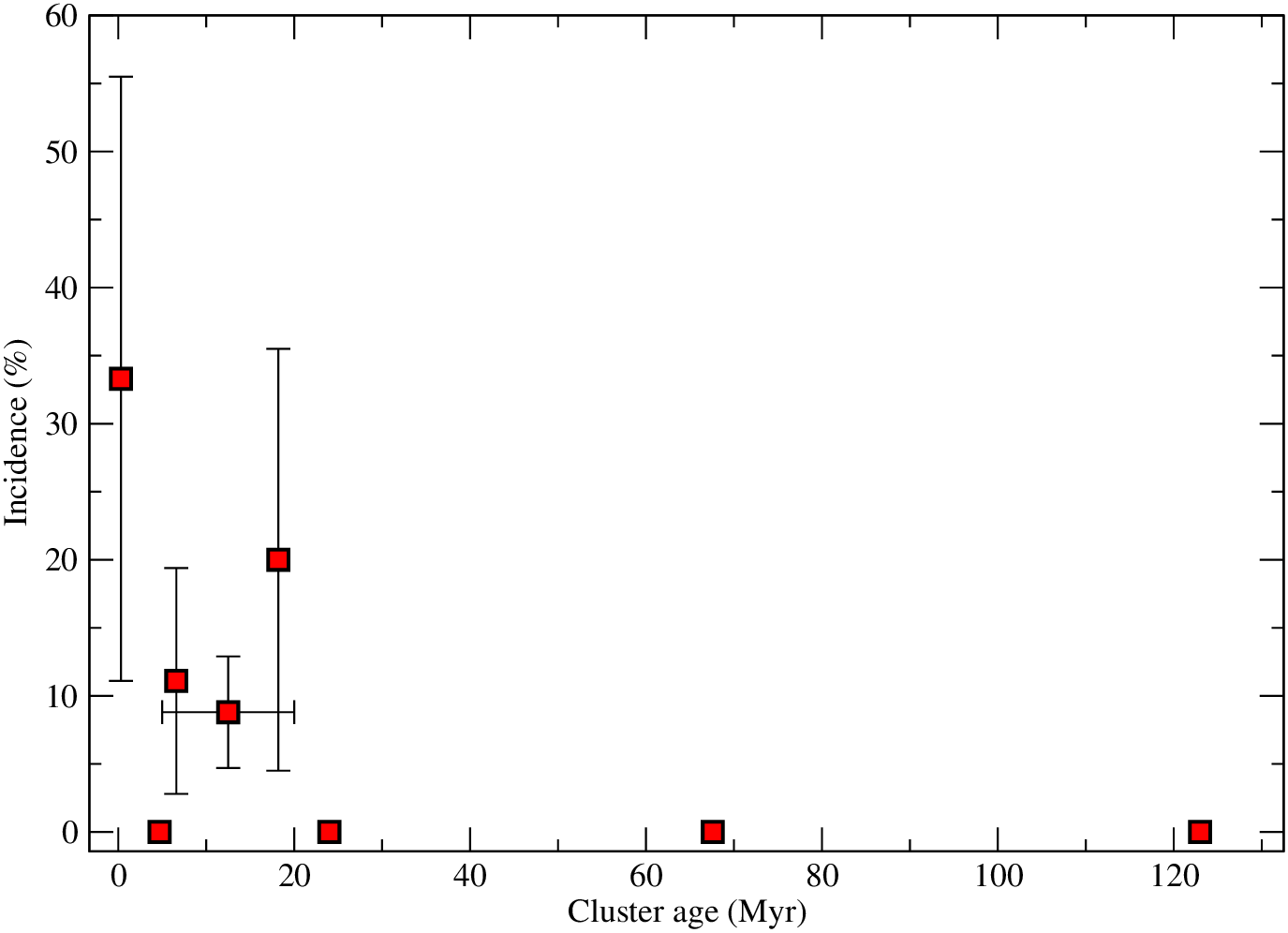}
\caption{{\em Left -}\ Magnetic incidence (fraction of magnetic stars as a function of all stars in the survey sample) versus spectral type. Error bars are inferred from counting statistics. No significant variation of incidence with spectral type is inferred. {\em Right -}\ Magnetic incidence versus age for a sample of open clusters observed in the context of the ESO HARPSpol LP, and including results for the ONC by Petit et al. (2008). }
\end{center}
\end{figure}

\section{Conclusion}

The MiMeS project provides a broad and robust survey of magnetism in bright Galactic stars with spectral types B9-O4. It establishes that the basic physics of magnetism in stellar radiative zones remains unchanged across 1.5 decades of stellar mass, from main sequence spectral type $\sim$F0 to O4. Papers reviewing the overall project, summarizing and analyzing the magnetic field results, and investigating the physical properties and systematics of various subsamples are currently in preparation, and will be submitted late 2013 through mid 2014.


\begin{thebibliography}{}

\bibitem[]{}E. Alecian, O. Kochukhov, C. Neiner, G.A. Wade et al., 2011, A\&A 536, 6
\bibitem[]{}M. Briquet, C. Neiner, B. Leroy, P.I. P\'apics, 2013, A\&A 557, 16
\bibitem[]{}J.-F. Donati, M. Semel, B.D. Carter, D.E. Rees, et al., 1997, MNRAS 291,658
\bibitem[]{}J. Grunhut, G.A. Wade, W. Marcolino, V. Petit et al., 2009, MNRAS 400, 94
\bibitem[]{}O. Kochukhov, V. Makaganiuk, N. Piskunov, 2010, A\&A 524, 5
\bibitem[]{}L. Mestel 1999, in Magnetic Fields Across the HR Diagram, ASP Conf. Proc. Vol. 248, 3
\bibitem[]{}C. Neiner, J. Grunhut, V. Petit, A. ud-Doula, A et al., 2012, MNRAS 426, 2738
\bibitem[]{}V. Petit, G.A. Wade, L. Drissen, T. Montmerle, E. Alecian, 2008, MNRAS 387, 23
\bibitem[]{}V. Petit \& G.A. Wade, 2012, MNRAS, 420, 73
\bibitem[]{}J. Silvester, O. Kochukhov \& G.A. Wade, 2013, MNRAS, submitted 
\bibitem[]{}M. Shultz, G.A. Wade, J. Grunhut, S. Bagnulo et al., ApJ 750, 2
\bibitem[]{}J. Sundqvist, A. ud-Doula, S. Owocki, R. Townsend, et al., 2012, MNRAS 423, 21
\bibitem[]{}G.A. Wade, J. Ma\' iz Apell\' aniz, F. Martins, V. Petit, et al., 2012, MNRAS 425, 1278



\end{thebibliography}
\end{document}